\documentclass[aps,twocolumn,longbibliography] {revtex4-1}

\usepackage{amsmath,graphicx,amsfonts}
\usepackage{mathrsfs}

\usepackage{times}
\usepackage[colorlinks=true,linkcolor=blue,urlcolor=blue,citecolor=blue,breaklinks=true]{hyperref}

\usepackage[utf8]{inputenc}
\DeclareUnicodeCharacter{03BC}{\mbox{$\mu$}}
\DeclareUnicodeCharacter{2009}{}

\DeclareUnicodeCharacter{03B4}{$\delta$}
\DeclareUnicodeCharacter{2009}{-}

\usepackage[english]{babel}

\makeatletter
\def\bbl@set@language#1{%
	\edef\languagename{%
		\ifnum\escapechar=\expandafter`\string#1\@empty
		\else\string#1\@empty\fi}%
	\@ifundefined{babel@language@alias@\languagename}{}{%
		\edef\languagename{\@nameuse{babel@language@alias@\languagename}}%
	}%
	\select@language{\languagename}%
	\expandafter\ifx\csname date\languagename\endcsname\relax\else
	\if@filesw
	\protected@write\@auxout{}{\string\select@language{\languagename}}%
	\bbl@for\bbl@tempa\BabelContentsFiles{%
		\addtocontents{\bbl@tempa}{\xstring\select@language{\languagename}}}%
	\bbl@usehooks{write}{}%
	\fi
	\fi}
\newcommand{\DeclareLanguageAlias}[2]{%
	\global\@namedef{babel@language@alias@#1}{#2}%
}
\makeatother

\DeclareLanguageAlias{en}{english}

\def\eq#1{Eq.~(\ref{#1})}

\def\L{\mathsf L}

\date\today
\begin{document}
	
\title{Analytical results for the capacitance of a circular plate capacitor}
\author{Benjamin Reichert}
\author{Zoran Ristivojevic}
\affiliation{Laboratoire de Physique Th\'{e}orique, Universit\'{e} de Toulouse, CNRS, UPS, 31062 Toulouse, France}
	
\begin{abstract}
We study the classic problem of the capacitance of a circular parallel plate capacitor. At small separations between the plates, it was initially considered in the 19th century by Kirchhoff who found the leading and the subleading term in the capacitance. Despite a large interest in the problem, almost 150 years  later, only the second subleading term has been found analytically. Using the recent advances in the asymptotic analysis of Fredholm integral equations of the second kind with finite support, here we study the one governing the circular capacitor, known as the Love equation. We find analytically many subleading terms in the capacitance at small separations. We also calculate the asymptotic expansion at large separations, thus providing two simple expressions which practically describe the capacitance at all distances. The approach described here could be used to find exact analytical expansions for the capacitance to an arbitrary number of terms in  regimes of both small and large separations. 
\end{abstract}

\maketitle

\section{Introduction}

Capacitance is one of the basic concepts in electrodynamics. For a capacitor, it  denotes the ratio between the charge on one of the plates and the potential difference between them. The capacitance purely depends on the geometry. The standard simplification in the textbooks is a parallel plate capacitor in a vacuum with the characteristic plate size much larger than their separation. In this case, the capacitance has the familiar form 
\begin{align}\label{eq:Cideal}
C=\epsilon_0 \frac{S}{\kappa}.
\end{align}
Here $S$ denotes the surface of the plates, $\kappa$ is their separation, while the constant $\epsilon_0$ is the vacuum permittivity. The expression (\ref{eq:Cideal}) should be understood only as a result valid in the limit $\kappa\to 0^+$ where the edge effects are neglected. 

The effects of the edge can be qualitatively understood through the example of a circular parallel plate capacitor. Let us consider such a system where the coaxial thin plates have unit radius and which are at the separation $\kappa\ll 1$. The capacitance of this system was initially studied by Kirchhoff \cite{kirchhoff_zur_1879} in 1877 who found 
\begin{align}\label{eq:Ckirchhoff}
\mathcal{C}(\kappa)=\frac{1}{4\kappa}+\frac{1}{4\pi}\left(\ln\frac{16\pi}{\kappa}-1\right)+ O(\kappa).
\end{align}
Here for convenience we have introduced $\mathcal{C}=C/{4\pi\epsilon_0}$, which we, somewhat loosely, also call the capacitance and use in the following. The first term in the expansion (\ref{eq:Ckirchhoff}) is in accordance with Eq.~(\ref{eq:Cideal}) and thus represents the idealized situation where the edge effects have been neglected. The remaining terms in Eq.~(\ref{eq:Ckirchhoff}) account for the leading correction that describes the edge effects. 

The problem of capacitance of a circular capacitor is a complicated one that has attracted considerable attention from researchers working in physics but also in mathematics. The nonrigorous derivation of Kirchhoff was first proven in the work of \citet{hutson_circular_1963} in 1963. Further efforts \cite{shaw_circulardisk_1970,leppington_capacity_1970,chew_microstrip_1982,tracy_ground_2016,paffuti_giampiero_numerical_2017} have resulted in the next term in the expansion  (\ref{eq:Ckirchhoff}) that is proportional to $\kappa$ (see below), as well as the general structure of the series in Eq.~(\ref{eq:Ckirchhoff}) where the coefficients are to be evaluated \cite{soibelman_asymptotics_1996}. We note that various numerical and analytical approximating schemes have been developed to find the capacitance \cite{wintle_edge_1985,hwang_last-passage_2006,norgren_capacitance_2009,pastore_numerical_2011,milovanovic_properties_2013,ristivojevic_conjectures_2019}.

On the technical level, the problem of evaluation of the capacitance of a circular capacitor turns out to be directly related to an integral equation of Fredholm type, known as the Love equation \cite{love_electrostatic_1949} in the potential theory literature \cite{sneddon_mixed_1966}. The explicit solution of that equation is not known currently, which is reflected in a quite small number of known terms in Eq.~(\ref{eq:Ckirchhoff}), which require significant efforts to be obtained. Interestingly, the same integral equation was derived later by \citet{lieb_exact_1963}, who were studying a seemingly unrelated quantum problem of bosons in one dimension with contact repulsion. The mathematical connection between the two problems was first noted and used in Ref.~\cite{gaudin_boundary_1971}. 

Recently, big progress has been achieved in understanding the way to solve the Fredholm-type integral equations of the second kind with finite support, which often appear when one studies integrable models and field theories \cite{volin_mass_2010,volin_quantum_2011,marino_exact_2019,marino_resurgence_2019}. Using those achievements, in this paper we analyze the Love equation and calculate analytically the small $\kappa$ asymptotics of the capacitance to a high order. We also systematically solve the Love equation at large $\kappa$ and then analytically calculate the capacitance in this regime. 

\section{The Love equation}

Our goal is to find the capacitance of a circular capacitor that consists of two thin coaxial conducting disks of unit radius at the separation $\kappa$. The disks are  held at equal potentials in the absolute value, $\pm V_0/2$, which guarantees equal charges on the two surfaces, $\pm Q$. By the definition, the capacitance is given by $C=Q/V_0$. To find that ratio, one should solve  the Laplace equation for the potential with appropriate boundary conditions for the potential on the disks. This could be done in an elementary way \cite{carlson_circular_1994}. The central quantity that determines the capacitance is encoded into the Love integral equation \cite{love_electrostatic_1949,sneddon_mixed_1966}
\begin{align}
\label{eq:Love}
f(x,\kappa)-\frac{\kappa}{\pi}\int_{-1}^{1} dy \frac{f(y,\kappa)}{\kappa^2+(y-x)^2}=1.
\end{align}
The latter equation determines the function $f(x,\kappa)$, which then enables one to express the capacitance $\mathcal{C}=C/4\pi\epsilon_0$ in the form
\begin{align}\label{eq:cap}
\mathcal{C}(\kappa)=\frac{1}{2\pi}\int_{-1}^{1}d x f(x,\kappa).
\end{align}
From this point the problem of capacitance becomes equivalent to the problem of solving the Love integral equation (\ref{eq:Love}). We will solve it in the regimes of small and large $\kappa$ and find the asymptotic expansions for $\mathcal{C}(\kappa)$ to a high order, which practically covers all the distances (see Fig.~\ref{fig:nx}).

\section{The formal solution of the Love equation at small $\kappa$}

In order to solve the Love equation (\ref{eq:Love}), we are using a method developed by \citet{volin_mass_2010,volin_quantum_2011}, which has recently been adapted to study the problem of one-dimensional gas of bosons with contact interaction \cite{marino_exact_2019}. Here we rederive the method and obtain a closed-form solution in terms of a system of linear equations. We apply the method to the problem of the circular plate capacitor to obtain its capacitance in the regime $\kappa\ll 1$ to, in principle, an arbitrary order. We also comment on the similarities of the present approach with the well-known old work of \citet{popov_theory_1977}.

As already noticed in Ref.~\cite{popov_theory_1977} (see also Refs.~\cite{hutson_circular_1963,lieb_exact_1963}), working with a perturbative expansion in small $\kappa$ of the function $f(x,\kappa)$ defined by Eq.~(\ref{eq:Love}) becomes problematic at higher orders in $\kappa$ since the corrections become increasingly more divergent near the end of the support of the function, rendering the perturbative expression for $f(x,\kappa)$ nonintegrable in the expression (\ref{eq:cap}). This is a serious problem which practically hinders the calculation of corrections to Kirchhoff's result (\ref{eq:Ckirchhoff}). To overcome this issue, it is convenient to deal with the resolvent $R(z)$ of the function $f(x,\kappa)$ defined by
\begin{align}
R(z)=\int_{-1}^{1} dx \frac{f(x)}{z-x}. \label{R}
\end{align}
Notice that for simplicity we omitted the argument $\kappa$ from $R(z)$ and $f(x)$. The resolvent is an analytic function in the complex plane except for $z\in [-1,1]$. Its discontinuity along $[-1,1]$ determines the function $f(x)$ on the same interval, since 
\begin{align}
f(x)=\dfrac{i}{2\pi}[R(x+i 0)-R(x-i 0)].\label{disc}
\end{align}
Here we have used the formula
\begin{align}
\delta(y)=\frac{i}{2\pi}\left(\frac{1}{y+i0}-\frac{1}{y-i0}\right).
\end{align}
We notice that $f(x)$ for $|x|>1$ can be found, e.g., from the integral equation (\ref{eq:Love}) by performing the integration once Eq.~(\ref{disc}) is used in the  integrand.

By making use of the resolvent, the integral equation (\ref{eq:Love}) is transformed into a difference equation
\begin{align}
\left[1-\mathcal D(\kappa)\right]\widetilde R(x+i0)-\left[1-\mathcal D(-\kappa)\right]\widetilde R(x-i0)=0,\label{diff}
\end{align}
where 
\begin{align}\label{eq:RRt}
\widetilde R(z)=R(z)-\frac{\pi z}{ \kappa}.
\end{align}
Here we introduced the shift operator $\mathcal D(\kappa)=e^{i \kappa \partial_z}$ acting as
\begin{align}
\mathcal D(\kappa) R(z)=R(z+i \kappa).
\end{align}
Instead of solving the integral equation (\ref{eq:Love}) we should now solve the alternative equation (\ref{diff}). This will be achieved in two steps \cite{volin_mass_2010,volin_quantum_2011}, first considering the bulk regime near the origin and then in the edge regime, which is in the vicinity of $x=\pm 1$. One finally matches the two solutions for the resolvent in the overlapping regime, as discussed below. We notice that a similar procedure was also applied in the study of \citet{popov_theory_1977}, who was solving the integral equation (\ref{eq:Love}) directly, rather than dealing with the resolvent.

\subsection{Bulk regime}
Let us start by studying the bulk regime given by the limit
\begin{align}
\kappa \to 0,\qquad z\text{ fixed}.
\end{align}
In this case, we assume the resolvent  in the form 
\cite{volin_mass_2010,volin_quantum_2011,marino_exact_2019}
\begin{align}
&\widetilde R_b(z)=-\dfrac{\pi\sqrt{z^2-1}}{\kappa} \notag\\
	&+\sum_{n,m=0}^\infty\sum_{k=0}^{n+m+1}c_{n,m,k}\kappa^{m+n} \dfrac{z^{\lambda_k}}{(z^2-1)^{n+1/2}}\ln^k\left(\dfrac{z-1}{z+1}\right), \label{Rt}
\end{align}
where $\lambda_k=[1-(-1)^k]/2$, while the subscript $b$ denotes the bulk solution.
The ansatz for the resolvent (\ref{Rt}) is proposed in Ref.~\cite{marino_exact_2019}. We showed that it is consistent with the ansatz for $f(x)$ that \citet{popov_theory_1977} used to solve Eq.~(\ref{eq:Love}) (see the discussion below). In Eq.~(\ref{Rt}), the coefficients $c_{n,m,k}$ are unknown polynomials, as it turns out, of $\ln \kappa$. Therefore, they only weakly depend on $\kappa$. It is important to note that the ansatz (\ref{Rt}) contains many terms that diverge when $z\to \pm 1$. Moreover, each subsequent term that has higher value of $n$ is more divergent from the preceding ones. These two issues imply that the ansatz for the resolvent is justified only for $|z\pm 1|\gg \kappa$. Therefore, the ansatz (\ref{Rt}) applies in the complex plain sufficiently outside the two circles  around the centers at $\pm1$ with a small characteristic radius $\kappa$. This explains the name bulk solution, opposite to the edge solution that is derived as a series expansion near $z=1$ (see below).

The coefficients $c_{n,m,k}$ of Eq.~(\ref{Rt}) should be chosen in such a way that Eq.~(\ref{diff}) is satisfied. Substituting the ansatz (\ref{Rt}) into Eq.~(\ref{diff}) and using the expression for the logarithm around the branch cut of the form $(0<x<1)$
\begin{align}
\dfrac{\ln^k\left(\dfrac{x\pm i0-1}{x\pm i0+1}\right)}{[(x\pm i0)^2-1]^{n+1/2}}=\mp i (-1)^n\dfrac{\left(\ln\dfrac{1-x}{1+x}\pm i \pi\right)^k}{(1-x^2)^{n+1/2}},\label{cont}
\end{align}
in the limit $\kappa\to 0$ and $x\to0$ one obtains the relations that $c_{n,m,k}$ should satisfy. At order $\kappa^{N+1}$ and $x^{2N^2}$, one can find all the coefficients $c_{n,m,k>0}$ for $n+m<N$. In other words, the coefficients in front of logarithms in Eq.~(\ref{Rt}) can be fixed. Such a procedure parallels the one of \citet{popov_theory_1977}, who was working with the ansatz for $f$. Substituting it into the integral equation (\ref{eq:Love}), Popov was able to fix the coefficients in front of logarithms order by order in small $\kappa$. However, the determination of the remaining coefficients, in our case $c_{n,m,0}$, is not possible using the ansatz for the bulk regime. Instead one must solve the problem near the edge and match the bulk with the edge solution.

\subsection{Connection between the resolvent and $f(x)$ and moments of $f(x)$}

Using the relation (\ref{disc}) and Eq.~(\ref{cont}), from the resolvent (\ref{eq:RRt}) we obtain the ansatz for $f(x)$ in the bulk:
\begin{align}
f(x)={}&\dfrac{\sqrt{1-x^2}}{\kappa} \notag\\
&+\dfrac{1}{\pi}\sum_{n,m=0}^\infty\sum_{k=0}^{n+m+1}c_{n,m,k}\kappa^{m+n} \dfrac{(-1)^n x^{\lambda_k}}{(1-x^2)^{n+1/2}}\notag\\
&\times \sum_{p=0}^k\dfrac{k!\lambda_{k-p+1}}{ p!(k-p)!}(i \pi)^{k-p}\ln^p\left(\dfrac{1-x}{1+x}\right). \label{fR}
\end{align}
Equation (\ref{fR}) is a good ansatz only away from the endpoints of the support of $f(x)$. Indeed, as pointed out by \citet{popov_theory_1977}, higher-order contributions in the perturbative expansion (\ref{fR}) at small $\kappa$ are more and more divergent near $x=\pm 1$. One can easily find the estimate for the perturbative expansion to break down by looking at the consecutive terms in powers of $\kappa$.
They are of the same order when $1-x^2\sim \kappa$, implying the bulk ansatz (\ref{fR}) is only good for $1-x^2\gg \kappa$, i.e., not too close to $x=\pm 1$ \cite{popov_theory_1977}. Such behavior of $f(x)$ leads to issues when one tries to calculate different moments of $x$, and in particular the zeroth moment that is proportional to the capacitance (\ref{eq:cap}). The formal expression obtained by substituting Eq.~(\ref{fR}) into the capacitance (\ref{eq:cap}) is divergent. One should thus find a way to treat this problem, since the capacitance is not expected to diverge at any finite $\kappa$. 

A simple solution of the latter problem involves the resolvent (\ref{R}) rather than the function $f(x)$  \cite{volin_quantum_2011,marino_exact_2019}. Let us first expand the resolvent at $x\to \infty$. 
From the definition  (\ref{R}), one expresses it in the form 
\begin{align}
R(x)=\sum_{n=0}^\infty T_n(\kappa) x^{-n-1}, \label{mom}
\end{align}
where the moments are defined as 
\begin{align}\label{eq:Tn}
T_n(\kappa)=\int_{-1}^{1}dx x^n f(x).
\end{align}
Since $R(x)=\widetilde R_b(x)+\pi x/\kappa$, by expanding $\widetilde R_b(x)$ of Eq.~(\ref{Rt}) near $x=\infty$ one obtains the moments $T_n(\kappa)$ as a function of the bulk coefficients $c_{n,m,k}$. In particular, the zeroth moment is given by $T_0(\kappa)=2\pi\mathcal{C}(\kappa)$, where the capacitance is
\begin{align}
\mathcal{C}(\kappa)=\dfrac{1}{4\kappa}+\dfrac{1}{2\pi}\sum_{m=0}^\infty (c_{0,m,0}-2c_{0,m,1})\kappa^m. \label{cap2}
\end{align}
Using $f(x)$ given by \eq{fR} makes the integral of $f(x)$ contained in the definition of $T_0$ (\ref{eq:Tn}) divergent due to the divergence of terms with $n>0$ as $|x|\to1$. However, it is very interesting to notice that $T_0(\kappa)$ only contains  the coefficients $c_{0,m,0}$ and $c_{0,m,1}$, i.e., it is determined by the terms with $n=0$ from the ansatz (\ref{fR}). Such truncated ansatz that contains only $n=0$ terms is actually integrable. We were able to explicitly calculate the integral from $-1$ to $1$ of the truncated $f(x)$. Using
\begin{gather}
\frac{1}{\pi}\int_{-1}^{1} dx \frac{1}{\sqrt{1-x^2}}\ln^p\left(\frac{1-x}{1+x}\right)=(1-\lambda_{p}) (i \pi)^p E_p,\\ 
\frac{1}{\pi}\int_{-1}^{1} dx \frac{x}{\sqrt{1-x^2}}\ln^p\left(\frac{1-x}{1+x}\right)=2i \lambda_{p} (i \pi)^p p E_{p-1}, 
\end{gather}
where $E_p$ is the Euler number, we performed the summation over $p$. We obtained that the only nonzero contribution is the one arising from $k=0$ or $k=1$, which leads to the right-hand side of Eq.~(\ref{cap2}) multiplied by $2\pi$.

The preceding discussion implies that the problem of the calculation of the capacitance [see Eqs.~(\ref{eq:cap}) and (\ref{cap2})] simply becomes a determination of the coefficients $c_{0,m,0}$ and $c_{0,m,1}$. As discussed below \eq{cont}, unlike $c_{0,m,1}$, one cannot obtain the coefficients $c_{0,m,0}$ only from the solution in the bulk. We therefore now consider the problem near the edge.

\subsection{Edge regime}

Let us now focus on the edge regime, which is defined by the limit
\begin{align}
\kappa\to 0,z\to 1,\qquad t=2\frac{z-1}{\kappa}\quad\text{fixed}. \label{edge}
\end{align}
The starting integral equation of the form of Eq.~(\ref{eq:Love}) in the edge regime at leading order in small $\kappa$ was solved by the Wiener-Hopf method in Refs.~\cite{hutson_circular_1963,popov_theory_1977,pustilnik_fate_2015}. However, here we need more terms of the expansion in $\kappa$, since the resolvent in the edge regime is needed to fix the unknown coefficients $c_{0,m,0}$. They will enable us to find the capacitance (\ref{cap2}) at higher orders in small $\kappa$. In order to find an ansatz, we use the  Laplace transform $\hat R(s)$ of $\widetilde R(z=1+\kappa t/2)$ defined by
\begin{align}\label{eq:LT}
\widetilde R(1+\kappa t/2)=\int_{0}^{\infty} ds e^{-s t}\hat R(s).
\end{align}
This enables us to write the difference equation (\ref{diff}) as an equation that holds for $s<0$ of the form \cite{volin_mass_2010,marino_exact_2019}
\begin{align}
\sin(s)\left[e^{i s}\hat R(s+i0)+e^{-i s}\hat R(s-i0)\right]=0.\label{diffL}
\end{align}
Equation (\ref{diffL}) imposes constraints on the form that  $\hat R(s)$ can have \cite{volin_mass_2010,volin_quantum_2011}, such as the condition that $\hat R(s)$ must be analytic everywhere except on the negative real axis at each order in small $\kappa$ expansion. The general solution of Eq.~(\ref{diffL}) is given by \cite{volin_mass_2010,marino_exact_2019}
\begin{align}
\hat R_e(s)={}&\frac{1}{\sqrt{\kappa} s^{3/2}}\exp\left(\dfrac{s}{\pi}\ln\dfrac{\pi e}{s} \right)\Gamma\left(\dfrac{s}{\pi}+1\right)\notag\\
	&\times \sum_{m=0}^\infty\sum_{n=0}^{\infty}Q_{n,m}\frac{\kappa^{m+n}}{s^n}. \label{Rs}
\end{align}
Here the index $e$ refers to the edge, while the coefficients $Q_{n,m}$ are unknown polynomials of $\ln \kappa$ to be determined using the matching procedure. $\Gamma(x)$ denotes the gamma function. We note that Eq.~(\ref{diffL}) does not uniquely determine the solution (\ref{Rs}). Namely, each half integer instead of $3/2$ in the first term $1/s^{3/2}$ would nullify Eq.~(\ref{diffL}). However, the lowest one is fixed after comparing Eq.~(\ref{Rs}) with the inverse Laplace transform of the leading-order term of \eq{Rt} evaluated in the edge regime (\ref{edge}) (see Appendix~\ref{appendix:ILT}), which is given by
\begin{align}
\widetilde R_b(z=1+\kappa t/2)=-\frac{\pi}{\sqrt{\kappa}}\sqrt{t}.
\end{align}
Its inverse Laplace transform is $\sqrt{\pi}/2\sqrt{\kappa} s^{3/2}$, which matches the leading-order term $\propto Q_{0,0}$ of Eq.~(\ref{Rs}) at small $s$. This also fixes $Q_{0,0}=\sqrt{\pi}/{2}$.

\subsection{Matching the bulk and the edge solutions}

Now that we have obtained general expressions for the resolvent in the bulk regime [\eq{Rt}] and in the edge regime [\eq{Rs}], let us match them in order to fix all the unknown coefficients $Q_{n,m}$ and $c_{n,m,k}$. In order to proceed, one needs to either perform a Laplace transform of \eq{Rs} or an inverse Laplace transform of \eq{Rt}. We choose to do the latter. The matching procedure therefore becomes equivalent to the problem of solving the equation
\begin{align}
\hat R_b(s)=\hat R_e(s), \label{RsRt}
\end{align} 
where $\hat R_b(s)$ stands for the inverse Laplace transform of the bulk solution (\ref{Rt}) evaluated in the edge regime (\ref{edge}). In other words [see~Eq.~(\ref{eq:LT})],
\begin{align}\label{eq:ILTR}
\hat R_b(s)=\mathscr{L}^{-1}[R_b(1+\kappa t/2)].
\end{align}
Here $\mathscr{L}^{-1}[\ldots]$ denotes the inverse Laplace transform. For a function $F(t)$, it is defined as
\begin{align}\label{eq:ILT}
\mathscr{L}^{-1}[F(t)]=\int_{\varepsilon-i\infty}^{\varepsilon+i\infty} \frac{dt}{2\pi i} e^{st} F(t).
\end{align}
Here $\varepsilon$ is an arbitrary positive constant chosen is such way that the contour of integration lies to the right of all singularities of $F(t)$.

Equation (\ref{RsRt}) will be solved order by order at small $\kappa$, and thus it is convenient to perform the inverse Laplace transform~(\ref{eq:ILTR}) on the expansion of $\widetilde R_b(1+\kappa t/2)$ in the limit $\kappa\to 0$.
As a consequence, we need to calculate $\mathscr{L}^{-1}[t^{m-1/2}\ln^n t]$ for $n\geq 0$ and at integer $m$. However, the inverse Laplace transform exists only for $m\leq 0$. To deal with this issue we use the equality $\ln^n A=\lim_{x\to0}\frac{\partial^n}{\partial x^n}e^{x \ln A}$ to obtain an analytic continuation  for $m> 0$. As a result, one has the following analytic continuation under the inverse Laplace transform 
\begin{align}
\mathscr{L}^{-1}[t^{m-1/2}\ln^n t]=\frac{1}{s^{m+1/2}}\left[\frac{e^{-x\ln s}}{\Gamma(1/2-m-x)}\right]^{(n)}_{x=0}.
\end{align} 
Here and in the following we use the notation 
\begin{align}
\lim_{x\to 0}\frac{\partial^n}{\partial x^n}A(x)= [A(x)]^{(n)}_{x=0}.
\end{align}

The evaluation of Eq.~(\ref{eq:ILTR}) is tedious. The main steps are given in Appendix \ref{appendix:ILT}. It yields
\begin{widetext}
\begin{gather}\label{eq:Rbilt}
\hat R_b(s)=-\sqrt{\frac{\pi}{4\kappa s}} \sum_{n=0}^\infty \frac{[(2n)!]^2}{(4^n n!)^3}\frac{2n+1}{2n-1} \frac{\kappa^n}{s^{n+1}}
+\frac{1}{\sqrt{\kappa s}} \sum_{m=0}^\infty\sum_{n=-m}^\infty\sum_{\ell=0}^{n+m+1}\frac{1}{\ell!} \kappa^m {s}^n\left(\ln\frac{\kappa}{4s}\right)^\ell  V_b(n,m,\ell),
\end{gather}
where 
\begin{align}\label{vbbbb}
V_b(n,m,\ell)=\sum_{j=\text{max}(0,-n)}^m\sum_{k=\ell}^{n+m+1}  \frac{(-1)^{j}k!}{4^j j! (k-\ell)!} c_{n+j,m-j,k} \left[ \frac{\Gamma(n+2j+x+1/2)-2j \lambda_k \Gamma(n+2j+x-1/2)}{\Gamma(n-x+1/2)\Gamma(n+j+x+1/2)}\right]^{(k-\ell)} _{x=0}.
\end{align}
We recall $\lambda_k=[1-(-1)^k]/2$. The expression (\ref{Rs}) we can also bring  to the form of Eq.~(\ref{eq:Rbilt}):
\begin{align}\label{eq:ReILT}
\hat R_e(s)={}\frac{1}{\sqrt{\kappa s}}\sum_{n=0}^\infty Q_{n,0}\frac{\kappa^n}{s^{n+1}}+\frac{1}{\sqrt{\kappa s}} \sum_{m=0}^\infty\sum_{n=-m}^\infty\sum_{\ell=0}^{n+m+1}\frac{1}{\ell!}\kappa^m {s}^n \left(\ln\frac{\kappa}{4s}\right)^\ell V_e(n,m,\ell),
\end{align}
where
\begin{gather}\label{veeee}
V_e(n,m,\ell)=\sum_{j=\text{max}(\ell,n+1)}^{m+n+1} \frac{1}{\pi^{j}(j-\ell)!} Q_{j-n-1,m+n+1-j}  \left[e^{x \ln\frac{4\pi e}{\kappa}}\Gamma(1+x) \right]^{(j-\ell)}_{x=0}.
\end{gather}
\end{widetext}
Equation (\ref{RsRt}) now becomes equivalent to
\begin{gather}
V_b(n,m,\ell)=V_e(n,m,\ell), \label{VV}
\end{gather}
provided 
\begin{gather}
Q_{n,0}=-\frac{\sqrt{\pi}}{2}\frac{[(2n)!]^2}{(4^n n!)^3} \frac{2n+1}{2n-1},\label{QQ}
\end{gather}
which is obtained from the terms that involve the single summation in Eqs.~(\ref{eq:Rbilt}) and (\ref{eq:ReILT}).

It is interesting to note that one can actually determine all the coefficients $c_{n,m,k}$ and $Q_{n,m}$ only from matching  the bulk and edge solutions, i.e., without first finding the coefficients $c_{n,m,k>0}$ using \eq{diff} in the bulk, as we discussed earlier. Indeed, in order to find the coefficients $c_{n_1,n_2,n_3}$ or $Q_{n_1,n_2}$, one needs to solve \eq{VV} for $n=n_1,m=n_2,\ell=n_3$ and $n=-n_1-1,m=n_1+n_2,\ell=0$, respectively. This leads to a recursive procedure which can be implemented on a computer \cite{marino_exact_2019, volin_mass_2010,volin_quantum_2011}. In Appendix~\ref{ilt} we illustrate the procedure through an example where we find the coefficients needed to obtain the first two corrections of the capacitance. Let us mention that we have (indirectly) verified Eqs.~(\ref{vbbbb}) and (\ref{veeee}) by finding the ground-state energy of the, related to Eq.~(\ref{eq:Love}), Lieb-Liniger model  (see Appendix~\ref{AppD}).

\section{Capacitance at small $\kappa$}

Now that all the coefficients $c_{n,m,k}$ can be systematically calculated, we can obtain the capacitance (\ref{cap2}) at $\kappa\ll 1$ to the desired order. We provide here the capacitance with five corrections:
\begin{align}\label{smalllkappa}
\mathcal{C}(\kappa)={}&\dfrac{1}{4\kappa}+\dfrac{\L-1}{4\pi}+\dfrac{\kappa}{16\pi^2}(\L^2-2)\notag\\
&+\dfrac{ \kappa^2}{64\pi^3}[2\L^2-1-3\zeta(3)]\notag\\
	&-\dfrac{\kappa^3}{384 \pi^4}\{2 \L^3-6 \L^2-3\L[1+3\zeta(3)]+24 \zeta(3)\}\notag\\
	&+\dfrac{\kappa^4}{12288 \pi^5}\{16\L^4-96\L^3-48\L^2[3\zeta(3)-1]\notag\\
	&+48\L[1+19\zeta(3)]+6-720\zeta(3)-405\zeta(5)\}\notag\\
	&+ O(\kappa^5),
\end{align}
where $\L=\ln(16\pi/\kappa)$, while $\zeta(n)$ denotes the zeta function. 
Our result is in agreement with the known expressions for the capacitance that was obtained at the linear order in $\kappa$ \cite{tracy_ground_2016,shaw_circulardisk_1970,chew_microstrip_1982}. However, the procedure described in this paper could be used to analytically find  an arbitrary number of terms, the only limitation being the computer time. In  Appendix~\ref{Cap} we provide the capacitance to the order $\kappa^7$.

\section{Capacitance at large $\kappa$}

\begin{figure}
	\centering
	\includegraphics[width=0.9\columnwidth]{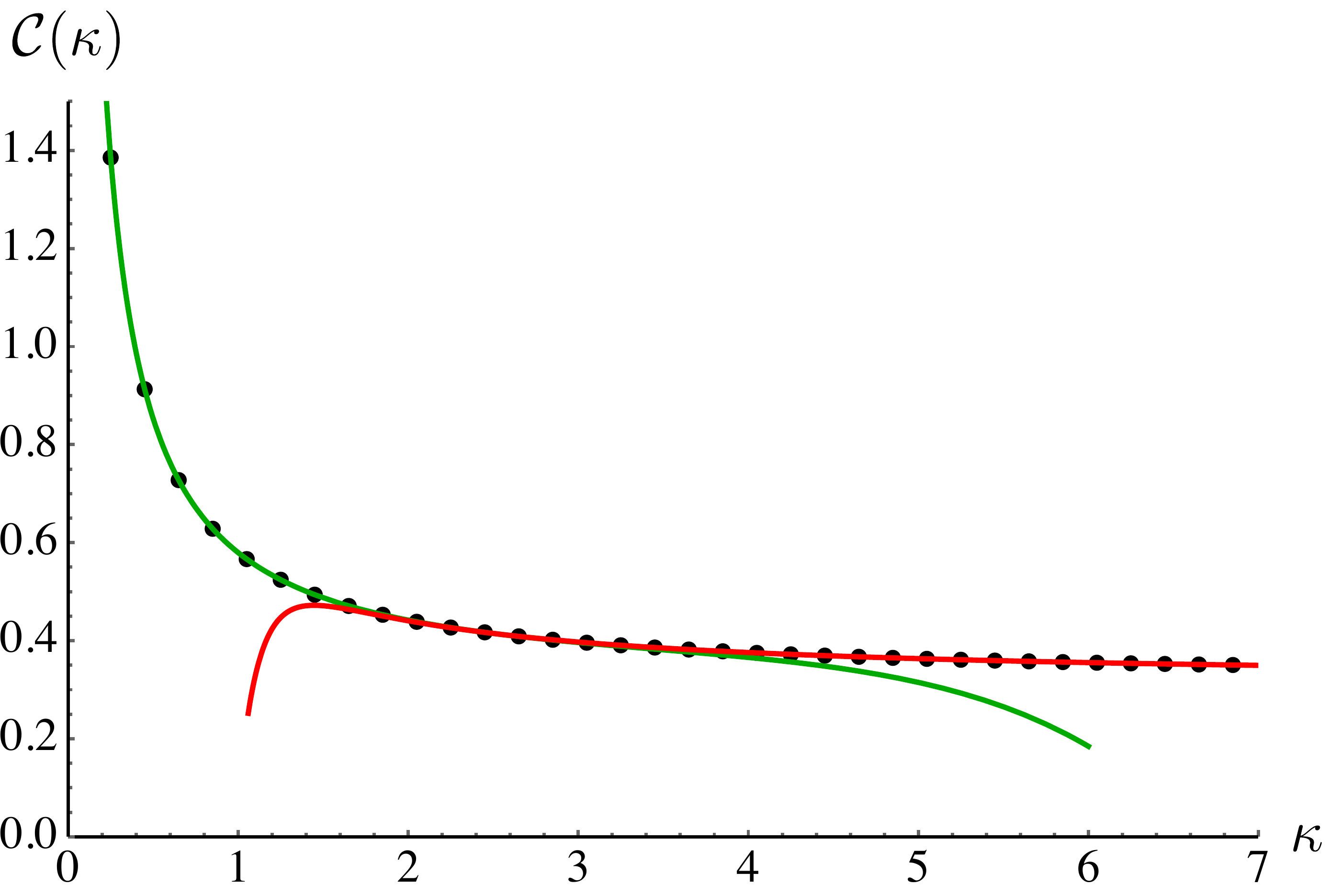}
	\caption{Capacitance $\mathcal C(\kappa)$ as a function of the distance between the plates $\kappa$. The black dots represent an approximation of the exact result obtained by solving numerically \eq{eq:Love}. The solid green line is the $\kappa\to0$ expansion to $O(\kappa^6)$ order given in Appendix~\ref{Cap}, while the solid red line is $\kappa\to\infty$ expansion (\ref{largekappa}).}
	\label{fig:nx}
\end{figure}

At $\kappa\gg 1$, the Love equation (\ref{eq:LT}) can be analytically solved in a systematic way using the expansion into orthogonal polynomials \cite{ristivojevic_excitation_2014}. There one assumes the solution in the form
\begin{align}\label{eq:fl}
f(x,\kappa)=\sum_{n=0}^\infty a_n(\kappa) P_n(x),
\end{align}  
where $P_n(x)$ is the Legendre polynomial. Since $f(x,\kappa)$ is an even function, only even polynomials are involved in Eq.~(\ref{eq:fl}). Using the orthogonality condition for the polynomials, Eq.~(\ref{eq:Love}) can be expressed as a system of equations for the coefficients $a_n(\kappa)$ of Eq.~(\ref{eq:fl}). It has the form
\begin{align}\label{system}
\frac{2 a_n(\kappa)}{2n+1}= {}&\sum_{m=0}^{M}\sum_{\ell=0}^{2m}\sum_{r=0}^{\ell} \frac{(-1)^{m+\ell}}{\pi} \binom{2m}{\ell} \frac{a_r(\kappa)}{\kappa^{2m+1}} F_{r}^\ell F_{n}^{2m-\ell}\notag\\
&+2\delta_{n,0},\quad M\to \infty.
\end{align}
Here $F_n^\ell=\int_{-1}^{1}dx\, x^\ell P_n(x)$, which is nonzero only for $\ell \geq n$, provided $\ell+n$ is an even integer. It is then given by
\begin{align}
F_n^\ell=\frac{2^{n+1} \ell! \left(\frac{\ell+n}{2}\right)!}{ (\ell+n+1)!\left(\frac{\ell-n}{2}\right)!}.
\end{align}
Using Eq.~(\ref{eq:fl}), the capacitance (\ref{eq:cap}) simply becomes
\begin{align}
\mathcal{C}(\kappa)=a_0(\kappa)/\pi.	
\end{align} 
Solving the system of equations (\ref{system}) at some fixed $M$, one obtains the capacitance with the precision $O(1/\kappa^{2M+2})$. For $M=3$ we obtain the asymptotic result
\begin{align}
\mathcal C(\kappa)={}&\dfrac{1}{\pi}+\dfrac{2}{\pi^2\kappa}+\dfrac{4}{\pi^3\kappa^2}-\dfrac{4(\pi^2-6)}{3\pi^4\kappa^3}-\dfrac{16(\pi^2-3)}{3\pi^5\kappa^4}\notag\\
&+\dfrac{16(2\pi^4-15\pi^2+30)}{15\pi^6\kappa^5}+\dfrac{32(\pi^4-4\pi^2+6)}{3\pi^7\kappa^6}\notag\\
	&-\dfrac{32(45\pi^6-371\pi^4+1050\pi^2-1260)}{315\pi^8\kappa^7}\notag\\
	&-\dfrac{64(128\pi^6-567\pi^4+1260\pi^2-1260)}{315\pi^9\kappa^8} \notag\\ &+O(1/\kappa^9). \label{largekappa}
\end{align}
In Fig.~\ref{fig:nx} we show the numerically evaluated capacitance that perfectly matches with our analytical formulas. 

\section{Summary}
In this paper we calculated the capacitance of the circular plate capacitor. To obtain the result in the regime of small plate separations, we used recently developed methods \cite{volin_mass_2010,volin_quantum_2011,marino_exact_2019,marino_resurgence_2019} to study the resulting integral equation, enabling us to obtain an asymptotic expansion  for the capacitance to a high order. We also found the series expansion of the capacitance at large plate separations. The obtained results (\ref{smalllkappa}) and (\ref{largekappa}) practically cover the whole range of distances (see Fig.~\ref{fig:nx}). If needed, they can be further calculated to an arbitrary precision. In Appendix \ref{AppD} we discuss the connection of our formulas to the problem of one-dimensional bosons. Let us finally notice that the edge effects on the capacitance play an important role for the measurement of the Casimir force \cite{wei_edge_2010}.

\section*{Acknowledgments}
This study has been partially supported through the EUR grant NanoX ANR-17-EURE-0009 in the framework of the ‘Programme des Investissements d’Avenir’.

\appendix

\begin{widetext}
\section{Inverse Laplace transform}\label{appendix:ILT}

Our aim is to calculate the inverse Laplace transform $\mathscr{L}^{-1}[\widetilde{R}_b]$ of the bulk resolvent (\ref{Rt}) in the edge regime (\ref{edge}), i.e., we substitute
$t={(z-1)}/{\beta\kappa}$ and account for the limit $\kappa\to 0$. Here $\beta=1/2$ is introduced for convenience. Equation (\ref{Rt}) in this case becomes
\begin{align}\label{Rbt_er}
\widetilde{R}_b=-\frac{\pi}{\kappa} \sqrt{\beta\kappa t(2+\beta\kappa t)} + \sum_{n,m,j=0}^{\infty} \sum_{k=0}^{m+n+1} \frac{c_{n,m,k}}{2^{n+j+1/2}} \kappa^{m+n} \binom{-n-1/2}{j} \frac{\partial^k}{\partial {x^{k}}}\left[ (\beta\kappa t)^{j-n-1/2}e^{x \ln\frac{\beta\kappa t}{2+\beta\kappa t}}(1+\lambda_k\beta\kappa t)\right]{\bigg{|}_{x=0}}. 
\end{align}
The inverse Laplace transform in the limit $\kappa\to 0$ of the first term in the right-hand side of Eq.~(\ref{Rbt_er}) is elementary. For the second term we use the asymptotic formula
\begin{align}\label{eq:ILT}
\mathscr{L}^{-1}\left[(\beta\kappa t)^{a} e^{x \ln\frac{\beta\kappa t}{2+\beta\kappa t}}\right]\simeq \frac{1}{s}\left(\frac{\beta\kappa}{s}\right)^a \left(\frac{\beta\kappa}{2s}\right)^x \frac{1}{\Gamma(-a-x)}\sum_{\ell=0}^{\infty} \frac{1}{\ell!} \frac{\Gamma(x+\ell)}{\Gamma(x)} \frac{\Gamma(1+x+a+\ell)}{\Gamma(1+x+a)} \left(\frac{\beta\kappa}{2s}\right)^\ell
\end{align}
in the limit $\kappa\to 0$, since in this case one can omit the exponentially small terms $\propto O(e^{-2s/\beta\kappa})$. The parameter $a$ in Eq.~(\ref{eq:ILT}) is a half-integer in our case. After transforming the multiple sum according to
\begin{align}
\sum_{m,n,j,\ell=0}^{\infty} F(m,\ell,j,n)=\sum_{m=0}^{\infty}\sum_{j=0}^{m} \sum_{n=-j}^{\infty} \sum_{\ell=0}^{j} F(m-j,\ell,j-\ell,n+j),
\end{align}  
we use the relation
\begin{align}
\sum_{\ell=0}^{j} \frac{1}{\ell\,!} \binom{-n-j-1/2}{j-\ell}\frac{\Gamma(x+\ell)}{\Gamma(x)} \frac{1}{\Gamma(1/2-x+n+\ell)\Gamma(1/2+x-n-\ell)}=\frac{(-1)^{n+j}\cos(\pi x)}{\pi} \binom{n+2j+x-1/2}{j}.
\end{align}	
It leads to
\begin{align}
\mathscr{L}^{-1}[\widetilde{R}_b]={}&-\frac{\pi\sqrt{2}}{\kappa} \sum_{j=0}^{\infty} \binom{1/2}{j} \frac{1}{2^j \Gamma(-j-1/2)} (\beta\kappa)^{j+1/2}s^{-j-3/2} +\sum_{m=0}^{\infty}\sum_{j=0}^{m}\sum_{n=-j}^{\infty}\sum_{k=0}^{m+n+1} \frac{(-1)^j}{2^{n+2j+1/2}}c_{n+j,m-j,k}\notag\\ &\times\kappa^{m-1/2}s^{n-1/2}\beta^{-n-1/2} \frac{\partial^k}{\partial x^k} \left\{\left(\frac{\beta\kappa}{2s}\right)^x \frac{\binom{n+2j+x-1/2}{j}}{\Gamma(1/2+n-x)}  \left[1+\lambda_k\frac{\beta\kappa}{s} (n-x-1/2)\right]\right\}\Bigg{|}_{x=0}.
\end{align}
The term proportional to $\lambda_k$ in the latter expression can be further transformed by changing the indices of summation, yielding
\begin{align}
\mathscr{L}^{-1}[\widetilde{R}_b]={}&-\frac{\pi\sqrt{2}}{\kappa} \sum_{j=0}^{\infty} \binom{1/2}{j} \frac{1}{2^j \Gamma(-j-1/2)} (\beta\kappa)^{j+1/2}s^{-j-3/2} +\sum_{m=0}^{\infty}\sum_{j=0}^{m}\sum_{n=-j}^{\infty}\sum_{k=0}^{m+n+1} \sum_{\ell=0}^{k} \frac{(-1)^{j}}{4^j (2\beta)^{n+1/2}j!}\binom{k}{\ell}\notag\\
&\times c_{n+j,m-j,k}\kappa^{m-1/2}s^{n-1/2} \left(\ln\frac{\beta\kappa}{2s}\right)^\ell \frac{\partial^{k-\ell}}{\partial x^{k-\ell}}
\left[\frac{\Gamma(n+2j+x+1/2)-2j\lambda_k \Gamma(n+2j+x-1/2)}  {\Gamma(n-x+1/2)\Gamma(n+j+x+1/2)}\right]\bigg{|}_{x=0}.
\end{align}
After changing the order of summations, one obtains the formula (\ref{eq:Rbilt}) of the main text.

\section{Detailed solution of the capacitance to $O(\kappa)$ order}
\label{ilt}

In this Appendix we explain the recursive procedure to find the capacitance \eqref{cap2} to $O(\kappa)$ order. We thus need to evaluate the coefficients $c_{0,0,0},c_{0,0,1},c_{0,1,0}$ and $c_{0,1,1}$. The first correction to the capacitance is given in terms of $c_{0,0,0}$ and $c_{0,0,1}$. The equation for the coefficient $c_{0,0,0}$ is obtained by solving $V_b(0,0,0)=V_e(0,0,0)$. It leads to
\begin{align}
c_{0,0,0}=c_{0,0,1}(\ln4+\gamma)+\dfrac{\ln(4\pi e/ \kappa)-\gamma}{2},
\end{align}
where $\gamma\approx 0.5772$ is the Euler–Mascheroni constant. Now we should find $c_{0,0,1}$ which is obtained by solving $V_b(0,0,1)=V_e(0,0,1)$, and trivially leads to
$c_{0,0,1}=1/2$. This fixes $c_{0,0,0}=[1+\ln(16 \pi /\kappa)]/2$. We now find the second correction to the capacitance, which is controlled by $c_{0,1,0}$ and $c_{0,1,1}$. We begin with $c_{0,1,0}$, which requires us to solve $V_b(0,1,0)=V_e(0,1,0)$. It yields
\begin{align}
c_{0,1,0}={}&\dfrac{3}{8}c_{1,0,0}+c_{0,1,1}(\ln4+\gamma)+c_{1,0,1}\dfrac{\ln4+2+\gamma}{8}+c_{0,1,2}\dfrac{\pi^2-2(\ln4+\gamma)^2}{2}\notag\\
&-c_{1,0,2}\dfrac{8+9\pi^2-2(\ln 64+3\gamma-2)^2}{48}+Q_{0,1}\frac{\ln(4\pi e /\kappa)-\gamma}{\sqrt\pi}-\frac{3[\ln(4\pi e /\kappa)-\gamma]^2}{64\pi}-\dfrac{\pi}{128},
\end{align}
where we used $Q_{1,0}=-3\sqrt{\pi}/32$ according to \eq{QQ}. Therefore, in order to find $c_{0,1,0}$, one needs six other coefficients: $c_{1,0,0}$, $c_{0,1,1}$, $c_{1,0,1}$, $c_{0,1,2}$, $c_{1,0,2}$, and $Q_{0,1}$. The relation for the first one is obtained by solving $V_b(1,0,0)=V_e(1,0,0)$ and reads
\begin{align}
c_{1,0,0}=c_{1,0,1}(\ln4+\gamma-2)-c_{1,0,2}\left[(\ln4+\gamma-2)^2-\dfrac{\pi^2-8}{2} \right]+\dfrac{[\ln(4\pi e /\kappa)-\gamma]^2}{8\pi}+\dfrac{\pi}{48}.
\end{align}
To obtain $c_{0,1,1}$, we solve $V_b(0,1,1)=V_e(0,1,1)$ and obtain
\begin{align}
c_{0,1,1}=-\dfrac{1}{8}c_{1,0,1}-c_{1,0,2}\dfrac{3\ln4+3\gamma-2}{4}+2c_{0,1,2}(\ln4+\gamma)+\dfrac{Q_{0,1}}{\sqrt\pi}-\dfrac{3[\ln(4\pi e/\kappa)-\gamma]}{32\pi}.
\end{align}
We now solve $V_b(1,0,1)=V_e(1,0,1)$ to find
\begin{align}
c_{1,0,1}=2c_{1,0,2}(\ln4+\gamma-2)+\dfrac{\ln(4\pi e /\kappa)-\gamma}{4\pi}.
\end{align}
We now find the last two coefficients with $\ell=2$ [see Eq.~(\ref{VV})]. Equation $V_b(0,1,2)=V_e(0,1,2)$ gives
\begin{align}
c_{0,1,2}=\dfrac{3}{8}c_{1,0,2}-\dfrac{3}{64 \pi}.
\end{align}
The coefficient $c_{1,0,2}$ is found from $V_b(1,0,2)=V_e(1,0,2)$ and trivially yields $c_{1,0,2}={1}/{8\pi}$. Finally, the remaining coefficient $Q_{0,1}$ is determined by the equation $V_b(-1,1,0)=V_e(-1,1,0)$ and produces
\begin{align}
Q_{0,1}=c_{0,0,1}\dfrac{3\ln4+3\gamma-4}{16\sqrt\pi}+\dfrac{c_{0,0,0}}{16\sqrt\pi}+\dfrac{3[\ln(4\pi e/\kappa)-\gamma]}{32\sqrt\pi}=\dfrac{\ln(16\pi/\kappa)}{8\sqrt\pi},
\end{align}
where we used the calculated values for $c_{0,0,0}$ and $c_{0,0,1}$. The above relations among the coefficients fix their values, yielding
\begin{align}
c_{0,1,0}=\dfrac{\ln^2(16\pi /\kappa)-2}{8\pi},\qquad c_{0,1,1}=0.
\end{align}
The capacitance (\ref{cap2}) now becomes
\begin{align}
C(\kappa)=\dfrac{1}{4\kappa}+\dfrac{1}{2\pi}(c_{0,0,0}-2c_{0,0,1})+\dfrac{\kappa}{2\pi}(c_{0,1,0}-2c_{0,1,1})=\dfrac{1}{4\kappa}+\dfrac{\L-1}{4\pi}+\dfrac{\kappa}{16\pi^2}(\L^2-2).
\end{align}
Here we recall $\L=\ln(16\pi/\kappa)$. The recursive procedure becomes cumbersome when done by manually; however it can be easily implemented on a computer enabling one to find the capacitance at higher orders in $\kappa$.

\section{Capacitance to $O(\kappa^7)$ order }
\label{Cap}
We express the capacitance in the form
\begin{align}
\mathcal{C}(\kappa)=\frac{1}{\pi}\sum_{j=-1}^\infty b_j \left(\frac{\kappa}{8\pi}\right)^j.
\end{align}
The first six coefficients are defined by Eq.~(\ref{smalllkappa}), while the following three are
\begin{align}
b_{5}={}&-\frac{64 \L^5}{5}+\frac{352
	\L^4}{3}+192 \L^3 [\zeta
(3)-1]-32 \L^2 [63 \zeta
(3)+1]+\L [4096 \zeta
(3)+1620 \zeta (5)+40]\notag\\
&+8
\left[-138 \zeta (3)+81
\zeta (3)^2-612 \zeta
(5)+1\right],\\
b_6={}&\frac{512 \L^6}{15}-\frac{1280
	\L^5}{3}+\L^4
\left[\frac{3712}{3}-768
\zeta
(3)\right]+\frac{256}{3}
\L^3 [135 \zeta (3)-7]-32
\L^2 [1276 \zeta (3)+405
\zeta (5)+14]\notag\\
&+16 \L
\left[2128 \zeta (3)-648
\zeta (3)^2+5301 \zeta
(5)+2\right]-28350 \zeta
(7)-90720 \zeta (5)+32928
\zeta (3)^2-2368 \zeta
(3)+\frac{64}{3},\\
b_7={}& -\frac{2048
	\L^7}{21}+\frac{70144
	\L^6}{45}+512 \L^5 [6 \zeta
(3)-13]+\L^4 [7936-60672
\zeta (3)]+\frac{128}{3}
\L^3 [7460 \zeta (3)+2025
\zeta (5)+14]\notag\\
&+96 \L^2
\left[-5248 \zeta (3)+1080
\zeta (3)^2-9375 \zeta
(5)-22\right]+\L
\biggl[183552 \zeta
(3)-700032 \zeta
(3)^2+2153664 \zeta
(5)+567000 \zeta
(7)\notag\\
&-\frac{896}{3}\biggr]+16
\left[622\zeta(3)+51128 \zeta
(3)^2-65835 \zeta (5)+27135 \zeta
(3)\zeta
(5)-115992 \zeta
(7)+2\right].
\end{align}

\section{Connection with the Lieb-Liniger model}\label{AppD}

In the context of one-dimensional quantum physics, the integral equation (\ref{eq:Love}) is well known to describe the ground-state properties of bosons with two-body contact repulsion $V(x)=c\delta(x)$, known as the Lieb-Liniger model \cite{lieb_exact_1963}. In particular, the ground-state energy per particle is given by 
$E_0=\frac{\hbar^2 n^2}{2m} e(\gamma)$, 
where $m$ is the mass of bosons, while $n$ is their density. The dimensionless parameter $\gamma=c/n$ can be connected to the solution of the integral equation (\ref{eq:Love}) as
\begin{align}\label{conn}
\gamma=2\pi\frac{\kappa}{T_0(\kappa)},
\end{align}
where $T_0(\kappa)$ is defined by Eq.~(\ref{eq:Tn}). The function $e(\gamma)$ is given by the expression \cite{lieb_exact_1963}
\begin{align}\label{eg}
e(\gamma)=4\pi^2\frac{T_2(\kappa)}{T_0(\kappa)^3},
\end{align}
where in the right-hand side one should express $\kappa$ as a function of $\gamma$ using their connection (\ref{conn}). Similarly to $T_0(\kappa)$ [cf.~Eq.~(\ref{cap2})], the second moment can also be obtained from the resolvent. It takes the form
\begin{align}
T_2(\kappa)=\dfrac{\pi}{8\kappa}+\sum_{m=0}^\infty \left[\left(\frac{1}{2}c_{0,m,0}-\frac{5}{3}c_{0,m,1}\right)\kappa^m +(4c_{0,m+1,2}+c_{1,m,0}-2c_{1,m,1})\kappa^{m+1} -8c_{0,m+2,3}\kappa^{m+2} \right]. \label{t2}
\end{align}
Once one evaluates Eq.~(\ref{eg}) at $\kappa\ll 1$, corresponding to the weak interaction between bosons $\gamma\ll 1$, one obtains the formula~(1) of Ref.~\cite{ristivojevic_conjectures_2019} to the order $O(\gamma^5)$ [as well as the equivalent expression (29) of Ref.~\cite{marino_exact_2019}, which is written to $O(\gamma^{9/2})$ order].   

\end{widetext}

%\bibliography{capabib}

%merlin.mbs apsrev4-1.bst 2010-07-25 4.21a (PWD, AO, DPC) hacked
%Control: key (0)
%Control: author (0) dotless jnrlst
%Control: editor formatted (1) identically to author
%Control: production of article title (0) allowed
%Control: page (1) range
%Control: year (0) verbatim
%Control: production of eprint (0) enabled
%

\end{document}